\documentclass[pra,twocolumn,floatfix,aps,superscriptaddress,showpacs,amsfonts]{revtex4}
\usepackage{graphicx}
\usepackage{array}
\usepackage{amsthm}
\usepackage{amsmath}
\usepackage{amssymb}
\usepackage{mathrsfs}

\begin{document}


\title{Ordering states of Tsallis relative $\alpha$-entropies of coherence}

\author{Fu-Gang Zhang}
\affiliation{School of Mathematics and Information Science, Shaanxi Normal University, Xi'an, 710119, China}
\affiliation{School of Mathematics and Statistics, HuangShan University, Huangshan, 245041, China}
\author{Lian-He Shao}
\affiliation{College of Computer Science, Shaanxi Normal University, Xi'an, 710119, China}

\author{Yu Luo}
\affiliation{College of Computer Science, Shaanxi Normal University, Xi'an, 710119, China}

\author{Yongming Li}
\email{liyongm@snnu.edu.cn}
\affiliation{School of Mathematics and Information Science, Shaanxi Normal University, Xi'an, 710119, China}
\affiliation{College of Computer Science, Shaanxi Normal University, Xi'an, 710119, China}

\date{\today}
\begin{abstract}
In this paper, we study the ordering states with Tsallis relative $\alpha$-entropies of coherence  and $l_{1}$ norm of coherence  for single-qubit states. We show that any Tsallis relative $\alpha$-entropies of coherence and $l_{1}$ norm of coherence give the same ordering for single-qubit pure states. However, they don't generate the same ordering  for  some high dimensional pure states, even though these states are pure.
We also consider three special Tsallis relative $\alpha$-entropies of coherence, such as $C_{1}$ ,$C_{2}$ and $C_{\frac{1}{2}}$, and show any one of these three measures and $C_{l_{1}}$ will not generate the same ordering for single-qubit mixed states.  Furthermore, we find that any two of these three special measures generate different ordering for single-qubit mixed states.


\end{abstract}
\eid{identifier}
\maketitle

\section{Introduction}
Quantum coherence is one of the most important physical resources in quantum mechanics, which can be used in quantum  optics~\cite{Scully97}, quantum information and quantum computation~\cite{Nielsen}, thermodynamics~\cite{Rodr13,berg14},
and low temperature thermodynamics~\cite{Horodecki13,Lostaglio15,Naras15}. Many efforts have been made in quantifying the coherence of quantum states~\cite{berg06}. The authors of Ref.~\cite{Baum14} proposed a rigorous framework to quantify coherence. The framework gave four conditions that any proper measure of the coherence must satisfy. Based on this framework, one can define suitable measures with respect to the prescribed orthonormal basis, including
the relative entropy of coherence and the $l_{1}$ norm of coherence~\cite{Baum14}. In addition, various other coherence measures were discussed~\cite{Swapan16,Yuan15,Shao15,Strel15,Napoli16,Zhang16,Yu16,Rastegin16}. Many further discussions about quantum coherence  were aroused~\cite{Liu16,Yao15,Du15,Cheng15,Bera15,Xi15,Winter16,Bromley15,Xu16,Yadin15,Bagan15,Chitambar16}.

Up to now, many different coherence measures have been proposed based on different physical contexts. For the same state, different values of coherence will be obtained by different coherence measures. In this case, a very important question appears, that is, whether these measures generate
the same ordering.
We say that two coherence measures  $C_{m}$ and $C_{n}$ generate the same ordering  if they satisfy the condition
\begin{equation}
C_{m}(\rho)\leq C_{m}(\sigma)\Leftrightarrow C_{n}(\rho)\leq C_{n}(\sigma)
\end{equation}
for any density operators $\rho$ and $\sigma$. Liu et al.~\cite{Liu16} showed that the relative entropy of coherence and the  $l_{1}$ norm of coherence  don't give the same ordering for some mixed states. The topics about ordering states were widely considered in entanglement measures~\cite{Ei99,Vi00,Zy02,Mi04,Wei03,Zi06} and quantum correlation measures~\cite{Sen03,Ali10,Lang11,Galve13,Okrasa12}.
Recently, the author of Ref.~\cite{Rastegin16} proposed  Tsallis relative $\alpha$-entropies of coherence. The author proved Tsallis relative $\alpha$-entropies of  coherence
satisfy the conditions of (C1),(C2a) and (C3). But the condition of (C2b), i.e. Monotonicity under incoherent selective measurements, seems to be more sophisticated. In fact, a counterexample  showed that Tsallis relative $\alpha$ entropies of coherence may violate the condition (C2b) in some situations. Whereas, these coherence measures satisfy a generalized monotonicity for average coherence under subselection based on measurement ~\cite{Rastegin16}.

In this paper, we study the ordering states with Tsallis relative $\alpha$-entropies of coherence and $l_{1}$ norm of coherence  for single-qubit states. First, we  show that any Tsallis relative $\alpha$-entropies of coherence and $l_{1}$ norm of coherence give the same ordering for single-qubit pure states. However, the condition (1) doesn't always satisfy for  high dimensional pure states.
Second, we consider three special Tsallis relative $\alpha$-entropies of coherence, such as $C_{1}$ ,$C_{2}$ and $C_{\frac{1}{2}}$, and show any one of these three measures and $C_{l_{1}}$ will not generate the same ordering for single-qubit mixed states.  Furthermore, we find that any two of these three special measures generate different ordering for single-qubit mixed states.

This paper  is organized as follows.  In Sec.~\ref{sec:pre}, we briefly review some notions related to Tsallis relative $\alpha$-entropies of coherence and $l_{1}$ norm of coherence.  In Sec.~\ref{sec:pure}, we show that  Tsallis relative $\alpha$-entropies of coherence and $l_{1}$ norm of coherence generate the same ordering for single-qubit pure states. In Sec.~\ref{sec:mixed}, we show that they may not generate the same ordering for some single-qubit mixed states, and we give some examples to show our results.  We summarize our results in Sec.~\ref{sec:conclusion}.

\section{preliminaries}\label{sec:pre}
In this section, we review some notions related to quantifying quantum coherence.
Considering a finite-dimensional Hilbert space $H$ with $d=dim(H)$.
Fix a basis $\{|i\rangle\}$, let $I$ be a set of incoherence states, which is of the form

\begin{equation}
\delta=\displaystyle\sum_{i=1}^{d}\delta_{i}|i\rangle\langle i|, \nonumber
\end{equation}
where $\delta_{i}\in[0,1]$, and $\sum_{i=1}^{d}\delta_{i}=1$.  Baumgratz et al.~\cite{Baum14} proposed that any proper measure of the coherence $C$ must
satisfy the following four conditions:

$(C1):$ $C(\rho)\geq 0$ and $C(\rho)= 0$ if and only if $\rho\in I$;

$(C2a):$ $C(\rho)\geq (\Phi(\rho))$, where $\Phi$ is any   incoherent completely positive and trace preserving maps;

$(C2b):$ $C(\rho)\geq \sum_{i}p_{i}C(\rho_{i})$, where $p_{i}=Tr(K_{i}\rho K^{\dag}_{i})$, $\rho_{i}=\frac{K_{i}\rho K^{\dag}_{i})}{Tr(K_{i}\rho K^{\dag}_{i})}$, for all ${K_{i}}$ with
$\sum_{i}K_{i}K^{\dag}_{i}=I$ and $K_{i}IK^{\dag}_{i}\subseteq I$.

$(C3):$ $\sum_{i} p_{i}C(\rho_{i})\geq C(p_{i}\rho_{i})$ for any ensemble $\{p_{i},\rho_{i}\}$.
It has been  shown that  $l_{1}$ norm of coherence  and relative entropy of coherence satisfy
these four conditions ~\cite{Baum14}.
$l_{1}$ norm of coherence~\cite{Baum14} is defined as

\begin{equation}
C_{l_{1}}(\rho)=\displaystyle\sum_{i\neq j}\mid \rho_{ij}\mid,
\end{equation}
here $\rho_{ij}$ are entries of $\rho$. The coherence measure defined
by the $l_{1}$ norm is based on the minimal distance of $\rho$
to the set of incoherent states $I$, $C_{D}(\rho) = min_{\delta\in I} D(\rho,\delta)$
with D being the $l_{1}$ norm, and there is $0 \leq C_{l_{1}}(\rho) \leq d-1$.
The upper bound is attained for the maximally coherent
state $\mid \varphi_{max}\rangle =\frac{1}{\sqrt{d}}\sum_{i=1}^{d}\mid i\rangle$.

Tsallis relative $\alpha$-entropies~\cite{Furuichi04,Hiai11} for the density matrices $\rho$ and $\delta$, denoted by $D_{\alpha}(\rho\|\delta)$, is defined as

\begin{equation}
D_{\alpha}(\rho\|\delta)=\frac{Tr(\rho^{\alpha}\delta^{1-\alpha})-1}{\alpha-1} \nonumber
\end{equation}
for $\alpha\in(0,1)\sqcup (1,\infty)$. $D_{\alpha}(\rho\|\delta)$ reduces to the von Neumann relative entropy when $\alpha\rightarrow1$~\cite{Furuichi04}, i.e.,

\begin{center}
$\displaystyle\lim_{\alpha\rightarrow 1} D_{\alpha}(\rho\|\delta)=S(\rho\|\delta)=Tr[\rho(\ln\rho-\ln\delta)]$.
\end{center}

 Tsallis relative $\alpha$-entropies of coherence~\cite{Rastegin16}, denoted by $C_{\alpha}(\rho)$, is defined as

\begin{equation}
C_{\alpha}(\rho)=\displaystyle\min_{\delta\in I}D_{\alpha}(\rho\|\delta). \nonumber
\end{equation}
$C_{\alpha}(\rho)$ reduces to relative entropy of coherence $C_{r}(\rho)$ when $\alpha\rightarrow1 $~\cite{Baum14}, i.e., $C_{1}(\rho)=C_{r}(\rho)=S(\rho_{diag})-S(\rho)$. The author of Ref.~\cite{Rastegin16}  proved that Tsallis relative $\alpha$-entropies of coherence   satisfy the conditions of (C1), (C2a) and (C3) for all $\alpha\in(0,2]$, but it may violate (C2b) in some situations. However,  these measures satisfy a generalized monotonicity for average coherence under subselection based on measurement as the following form~\cite{Rastegin16}.

 For all $\alpha\in(0,2]$, Tsallis relative $\alpha$-entropies of coherence $C_{\alpha}(\rho)$ satisfy

\begin{equation}
\sum_{i}p_{i}^{\alpha}q_{i}^{1-\alpha}C_{\alpha}(\rho_{i})\leq C_{\alpha}(\rho)
\end{equation}
where  $p_{i}=Tr(K_{i}\rho K^{\dag}_{i})$, $q_{i}=Tr(K_{i}\delta_{\rho} K^{\dag}_{i})$,
and $\rho_{i}=\frac{K_{i}\rho K^{\dag}_{i}}{p_{i}}$.

A .E. Rastegin ~\cite{Rastegin16} gave an elegant mathematical analytical  expression of Tsallis relative $\alpha$-entropies of coherence.
For all $\alpha\geq 0$ and $ \alpha\neq 1$, the  Tsallis relative $\alpha$-entropies of coherence $C_{\alpha}(\rho)$, for a state $\rho$, can be expressed as

\begin{equation}
C_{\alpha}(\rho)=\frac{1}{\alpha-1}\{r^{\alpha}-1\}
\end{equation}
where $r=\sum_{i}\langle i|\rho^{\alpha}|i\rangle^{\frac{1}{\alpha}}$. For the given $\rho$ and $\alpha$, based on this coherence measure, the nearest incoherence state from $\rho$ is the state

\begin{equation}
\delta_{\rho}=\frac{1}{r}\sum_{i}\langle i|\rho^\alpha|i\rangle\}^\frac{1}{\alpha}|i\rangle\langle i|. \nonumber
\end{equation}

 Considering an interesting case $\alpha=2$, we get

\begin{equation}
C_{2}(\rho)=(\sum_{j}\sqrt{\sum_{i}|\rho_{i,j}|^{2}})^2-1
\end{equation}
where $\rho_{i,j}=\langle i|\rho|j\rangle$. $C_{2}$ is a function of squared module $|\rho_{i,j}|^{2}$, we should  distinguish it from $l_{2}$ norm of coherence $C_{l_{2}}$.
$C_{l_{2}}$ is defined as

\begin{equation}
C_{l_{2}}(\rho)=\displaystyle\sum_{i\neq j}\mid \rho_{ij}\mid^2. \nonumber
\end{equation}
It has been shown that  $C_{l_{2}}$ doesn't satisfy the condition (C2b)~\cite{Baum14}. Although $C_{2}$ also violates the condition (C2b),  but it obeys a generalized monotonicity property Eq. (3)~\cite{Rastegin16}.

\section{ordering states with $C_{\alpha}$ and $C_{l_{1}}$ for single-qubit pure states}\label{sec:pure}
 In this section,
we  show that Tsallis relative $\alpha$-entropies of coherence and $l_{1}$ norm of
coherence generate the same ordering for single-qubit pure states.

Let $\mid \psi\rangle=\sqrt{p}\mid 0\rangle + e^{i\varphi}\sqrt{1-p}\mid 1\rangle$ be a single-qubit pure state, where $p\in[0,1]$. It is easy to calculate that $l_{1}$ norm  of coherence of $\mid \psi\rangle$ is equal to
   $C_{l_{1}}=2\sqrt{p(1-p)}$.
 Tsallis relative $\alpha$-entropies of coherence  is equal to $C_{\alpha}=\frac{1}{\alpha-1}\{r^\alpha-1\}$, where $r=p^{\frac{1}{\alpha}}+(1-p)^{\frac{1}{\alpha}}$.
So we have the following proposition.

 \textbf{Proposition 1}:
(1) $C_{l_{1}}$ is an increasing function for $p\leq \frac{1}{2}$,
 and it is a decreasing function for $p\geq \frac{1}{2}$.

(2) $C_{\alpha}$ is an increasing function for $p\leq \frac{1}{2}$,
 and it  is a decreasing function for $p\geq \frac{1}{2}$.



Proof: (1) It is clear that $C_{l_{1}}=2\sqrt{p(1-p)}$ is an increasing function for $p\leq \frac{1}{2}$, and  is a decreasing function for $p\geq \frac{1}{2}$.

(2) We first consider the derivation of $r$ with respect to $p$. It is obvious that

\begin{center}
$\frac{\partial r}{\partial p}=\frac{1}{\alpha}[p^\frac{1-\alpha}{\alpha}+(1-p)^\frac{1-\alpha}{\alpha}]$
$\left\{
\begin{array}
{cc}> 0,& \quad   \alpha <1 ,  p> \frac{1}{2},\\
<0, &\quad  \alpha <1,  p < \frac{1}{2}\\
<0,& \quad  \alpha >1,  p> \frac{1}{2},\\
>0,& \quad  \alpha >1 , p< \frac{1}{2}.
\end{array}
\right. $
\end{center}

So we can know

\begin{center}
$\frac{\partial C_{\alpha}}{\partial p}=\frac{\alpha}{\alpha-1}r^{\alpha-1}\frac{\partial r}{\partial p}$
$\left\{
\begin{array}
{cc}< 0,& \quad   \alpha <1 ,  p> \frac{1}{2},\\
>0, &\quad  \alpha <1,  p < \frac{1}{2}\\
<0,& \quad  \alpha >1,  p> \frac{1}{2},\\
>0,& \quad  \alpha >1 , p< \frac{1}{2}.
\end{array}
\right. $
\end{center}

Therefore, $C_{\alpha}$ is an increasing function for $p\leq \frac{1}{2}$,
 and is a decreasing function for $p\geq \frac{1}{2}$.

By the above proposition, we can show that $C_{\alpha}$ and $C_{l_{1}}$ give the same ordering
for single-qubit pure states. Let $\mid \psi\rangle=\sqrt{p}\mid 0\rangle + \sqrt{1-p}\mid 1\rangle$ and $\mid \varphi\rangle=\sqrt{q}\mid 0\rangle + \sqrt{1-q}\mid 1\rangle$ be two
single-qubit pure states. The following result can be obtained.

\textbf{Result 1}: $C_{\alpha}(\mid \psi\rangle)\leq C_{\alpha}(\mid \varphi\rangle)$ if and only if
$C_{l_{1}}(\mid \psi\rangle)\leq C_{l_{1}}(\mid \varphi\rangle)$.

Proof: It is easy to know $C_{\alpha}(p)=C_{\alpha}(1-p)$, $C_{l_{1}}(p)=C_{l_{1}}(1-p)$.
Without loss of generality, we can set $p,q\leq\frac{1}{2}$. In line with proposition 1, we have $C_{\alpha}(\mid \psi\rangle)\leq C_{\alpha}(\mid \varphi\rangle)$ if and only if $p\leq q$, and $p\leq q$ if and only if $C_{l_{1}}(\mid \psi\rangle)\leq C_{l_{1}}(\mid \varphi\rangle)$. Therefore, $C_{\alpha}(\mid \psi\rangle)\leq C_{\alpha}(\mid \varphi\rangle)$ if and only if
$C_{l_{1}}(\mid \psi\rangle)\leq C_{l_{1}}(\mid \varphi\rangle)$.


Result 1 shows, for any $\alpha \in (0,2]$, $C_{\alpha}$ and $C_{l_{1}}$ generate the same ordering for single-qubit pure states. Moreover, for any two $\alpha_{1}, \alpha_{2} \in (0,2]$, $C_{\alpha_{1}}$ and  $C_{\alpha_{2}}$
also generate the same ordering for single-qubit pure states. Some explicit examples as Fig. 1 can intuitively show our conclusion.

\begin{center}
\vspace{-0.2cm}
\includegraphics [scale=0.25]{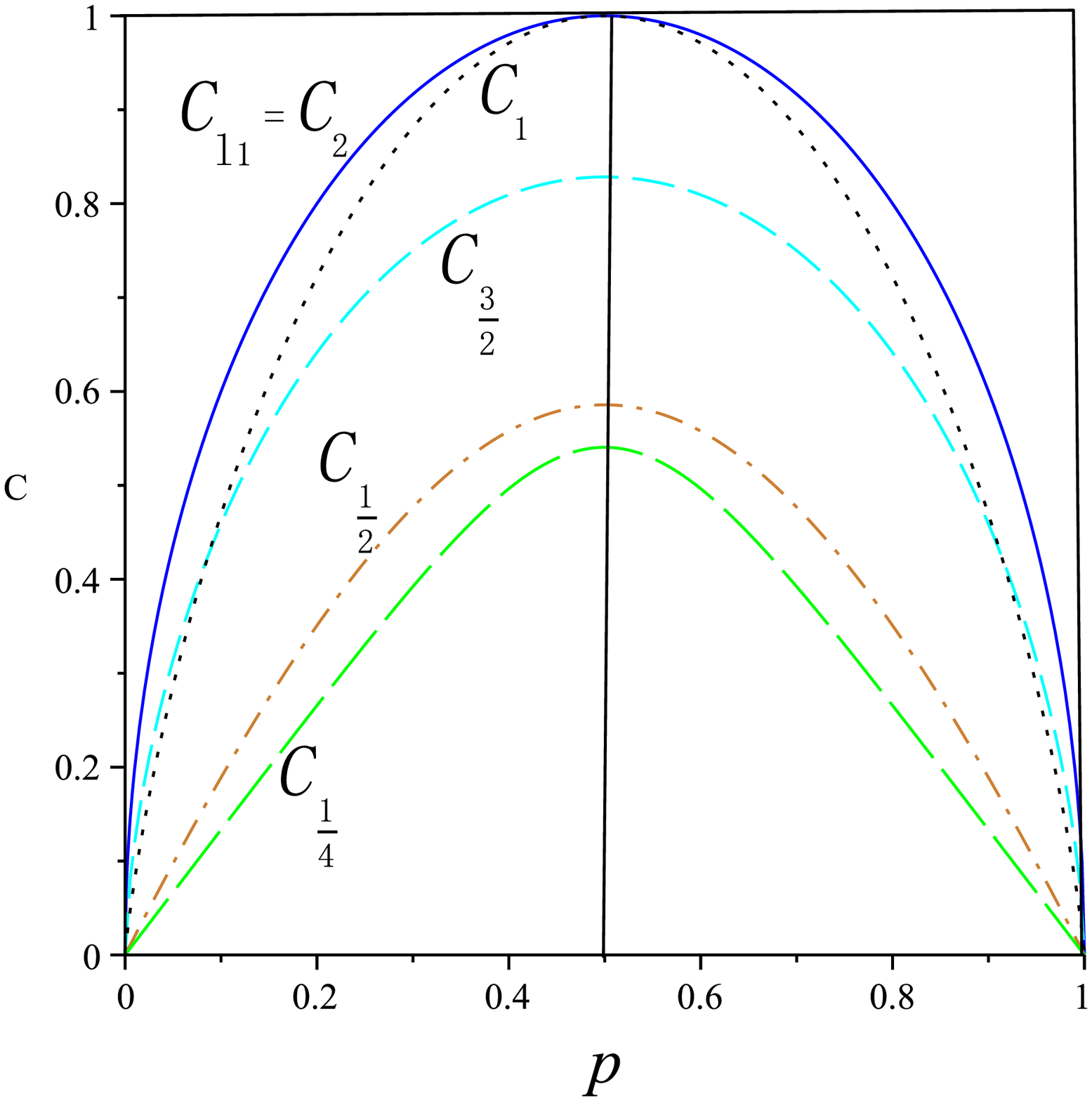}
\vspace{-0.2cm}
\center{Fig. 1.  Tsallis relative $\alpha$-entropies of coherence versus $l_{1}$ norm of
coherence for single-qubit pure states.}
\end{center}

It is worth noting that result 1 is only effective for single-qubit pure states. We find this result may be
invalid for states in high dimensional systems,  even though these states are pure. We give
a counterexample. Two qutrit pure states~\cite{Liu16} are given as follow,
\begin{center}
 $\mid\psi_{1}\rangle=\sqrt{\frac{12}{25}}\mid0\rangle+\sqrt{\frac{12}{25}}\mid 1\rangle+\sqrt{\frac{12}{25}}\mid 2\rangle$,
  $\mid\psi_{2}\rangle=\sqrt{\frac{7}{10}}\mid0\rangle+\sqrt{\frac{2}{10}}\mid 1\rangle+\sqrt{\frac{1}{10}}\mid 2\rangle$.
\end{center}

  It is easy to calculate that $C_{l_{1}}(\mid \psi_{1}\rangle)=1.5143$,
$C_{\frac{1}{2}}(\mid\psi_{1}\rangle)=0.6400$,  $C_{l_{1}}(\mid\psi_{2}\rangle)=1.5603$
$C_{\frac{1}{2}}(\mid\psi_{2}\rangle)=0.5303$. It is clear that $C_{l_{1}}(\mid\psi_{1}\rangle)< C_{l_{1}}(\mid\psi_{2}\rangle)$, and $C_{\frac{1}{2}}(\mid\psi_{1}\rangle)> C_{\frac{1}{2}}(\mid\psi_{2}\rangle)$.
So we  know that $C_{l_{1}}$ and $C_{\frac{1}{2}}$
 generate different ordering for single-qutrit pure states
  $\mid\psi_{1}\rangle$ and $\mid\psi_{2}\rangle$.

\section{ordering states with $C_{\alpha}$ and $C_{l_{1}}$ for single-qubit mixed states}\label{sec:mixed}

We consider ordering states with  $C_{\alpha}$ and $C_{l_{1}}$ for single-qubit mixed  states. Any single-qubit state $\rho$ can be written as~\cite{Nielsen}

\begin{equation}
\rho(x,y,z)=\left[ \begin {array}{cc} \frac{1+z}{2}&\frac{x-iy}{2}\\ \noalign{\medskip}\frac{x+iy}{2}&\frac{1-z}{2}\end {array} \right] \nonumber
\end{equation}
with $x^2+y^2+z^2\leq 1$. By a diagonal and unitary matrix~\cite{Liu16}, $\rho(x,y,z)$  can be transformed into a state with the form

\begin{equation}
\rho(t,z)=\left[ \begin {array}{cc} \frac{1+z}{2}&\frac{t}{2}\\ \noalign{\medskip}\frac{t}{2}&\frac{1-z}{2}\end {array} \right]
\end{equation}
with $t^2+z^2\leq 1$.
It has been shown that $\rho(x,y,z)$ and $\rho(t,z)$  can be transformed into each other by
an incoherent operation. Therefore, we only need to consider  states with the form $\rho(t,z)$. Next, we show $C_{\alpha}$ and $C_{l_{1}}$ generate the different ordering for some
single-qubit mixed states with the form Eq. (6).
Based on the Eq. (2), we get the $l_{1}$ norm of coherence of $\rho(t,z)$, $C_{l_{1}}(\rho(t,z))=t$.
Because the expression of $C_{\alpha}$ is sophisticated   for  any $\alpha\in(0,2]$,
so we only consider three special Tsallis relative $\alpha$-entropies of coherence, $C_{2}$, $C_{1}$ and $C_{\frac{1}{2}}$.

By substituting Eq. (6) into Eq. (5),  we obtain the expression of $C_{2}$.

 \begin{equation}
C_{2}(\rho)=r_{2}^2-1,
\end{equation}

where
 \begin{equation}
r_{2}=\frac{1}{2}\sqrt{(1+z)^2+t^2}+\frac{1}{2}\sqrt{(1-z)^2+t^2}.
\end{equation}

The authors of Ref.~\cite{Liu16}  considered the ordering states with the relative entropy of coherence $C_{r}$
 and the  $l_{1}$ norm of coherence $C_{l_{1}}$, and  obtained many useful results.
In this section, we see $C_{r}$ as a special $C_{\alpha}$ of $\alpha=1$. We  again discuss this question from our perspective.
 For $\alpha\rightarrow 1$,  Tsallis  $\alpha$-relative entropies reduce to the von Neumann relative entropy~\cite{Furuichi04}
\begin{equation}
\displaystyle\lim_{\alpha\rightarrow 1} D_{\alpha}(\rho\|\sigma)=S(\rho\|\delta)=Tr[\rho(\ln\rho-\ln\delta)]. \nonumber
\end{equation}
Thus $C_{\alpha}(\rho)$ reduce to $C_{r}(\rho)$.
We can denote

\begin{equation}
\begin{aligned}
C_{1}(\rho)&=C_{r}(\rho)=S(\rho_{diag})-S(\rho)\\
  &=\frac{1+\sqrt{t^2+z^2}}{2}\ln\frac{1+\sqrt{t^2+z^2}}{2}\\
 &+\frac{1-\sqrt{t^2+z^2}}{2}\ln\frac{1-\sqrt{t^2+z^2}}{2}\\
 &-\frac{1+z}{2}\ln\frac{1+z}{2}-\frac{1-z}{2}\ln\frac{1-z}{2}.
 \end{aligned}
\end{equation}

For $\alpha=\frac{1}{2}$, in order to calculate  $C_{\frac{1}{2}}$ of a mixed state $\rho(t,z)$ with the form Eq. (6), we need  eigenvalues and eigenvectors of this state.
With an easy calculation, we can obtain the eigenvalues of  $\rho(t,z)$,

\begin{center}
$\lambda_{1}=\frac{1+\sqrt{t^2+z^2}}{2}$, $\lambda_{2}=\frac{1-\sqrt{t^2+z^2}}{2}$.
\end{center}

Their norm eigenvectors are

\begin{center}
$|\lambda_{1}\rangle=\left[\frac{t}{(\sqrt{t^2+z^2}-z)^\frac{1}{2}*2\sqrt{t^2+z^2})^\frac{1}{2}},
\frac{(\sqrt{t^2+z^2}-z)^\frac{1}{2}}{(2\sqrt{t^2+z^2})^\frac{1}{2}}\right]^{T}$,
\end{center}

\begin{center}
$|\lambda_{2}\rangle=\left[\frac{-t}{(\sqrt{t^2+z^2}+z)^\frac{1}{2}*2\sqrt{t^2+z^2})^\frac{1}{2}},
\frac{(\sqrt{t^2+z^2}+z)^\frac{1}{2}}{(2\sqrt{t^2+z^2})^\frac{1}{2}}\right]^{T}.$
\end{center}

Substituting its eigenvalues and eigenvectors into Eq. (4), the expression of $C_{\frac{1}{2}}(\rho)$  can be given as:

\begin{equation}
    C_{\frac{1}{2}}(\rho)=-2\big(r_{\frac{1}{2}}^{\frac{1}{2}}-1\big),
\end{equation}
 $r_{\frac{1}{2}}$   is expressed as:

\[
\begin{aligned}
 r_{\frac{1}{2}}&=\Big[(\frac{1+\sqrt{t^2+z^2}}{2})^\frac{1}{2}\frac{\sqrt{t^2+z^2}+z}{2\sqrt{t^2+z^2}}\\
&+(\frac{1-\sqrt{t^2+z^2}}{2})^\frac{1}{2}\frac{\sqrt{t^2+z^2}-z}{2\sqrt{t^2+z^2}}\Big]^2 \\
 &+\Big[(\frac{1+\sqrt{t^2+z^2}}{2})^\frac{1}{2}\frac{\sqrt{t^2+z^2}-z}{2\sqrt{t^2+z^2}} \\
 &+(\frac{1-\sqrt{t^2+z^2}}{2})^\frac{1}{2}\frac{\sqrt{t^2+z^2}+z}{2\sqrt{t^2+z^2}}\Big]^2.
  \end{aligned}
\]

In the following, let us consider the monotonicity of expressions of these three coherence measures with respect to  variable $z$.

\textbf{Proposition 2}:
For a fixed value $t$, $C_{\alpha}(\rho(t,z))$ is an increasing function with respect to  $z$  for $0 \leq z\leq \sqrt{1-t^2}$, and it is a  decreasing function with respect to  $z$ for $-\sqrt{1-t^2}\leq z \leq 0$. i.e. $ \frac{\partial C_{\alpha}(\rho(t,z))}{\partial z}\geq 0$,  for  $0 \leq z\leq \sqrt{1-t^2}$, and  $ \frac{\partial C_{\alpha}(\rho(t,z))}{\partial z}\leq 0$, for $-\sqrt{1-t^2}\leq z \leq 0$, where $\alpha=2,1,\frac{1}{2}$ and $\rho(t,z)$ is single-qubit mixed state with the form Eq. (6).

Proof: Through analyzing the expression of $ C_{\alpha}(\rho)$,
we find  $C_{\alpha}(\rho(t,z))= C_{\alpha}(\rho(t,-z))$, where $\alpha=2,1,\frac{1}{2}$, $-\sqrt{1-t^2}\leq z\leq \sqrt{1-t^2}$.
Thus, we only need to show  that $ C_{\alpha}(\rho)$ is an increasing function for $0\leq z\leq \sqrt{1-t^2}$.

(1) We consider the derivation of $C_{2}(\rho)$ related to $z$.

 \begin{equation}
\frac{\partial C_{2}(\rho)}{\partial z}=r.\Big[\frac{1+z}{\sqrt{t^2+(1+z)^2}}-\frac{1-z}{\sqrt{t^2+(1-z)^2}}\Big]. \nonumber
\end{equation}

It is obvious  that $\frac{\partial C_{2}(\rho)}{\partial z}\geq 0$ for $0\leq z\leq \sqrt{1-t^2}$.

(2) For $0\leq z\leq \sqrt{1-t^2}$, we consider the derivation of $C_{\frac{1}{2}}(\rho)$ related to $z$,

\begin{equation}
    \frac{\partial C_{1}(\rho)}{\partial z}=\frac{1}{2}\ln\frac{1-z}{1+z}+\frac{z}{2\sqrt{t^2+z^2}}
   \ln\frac{1+\sqrt{z^2+t^2}}{1-\sqrt{z^2+t^2}}\geq 0. \nonumber
\end{equation}

 If   $f(x)=\frac{1}{x}\ln\frac{1+x}{1-x}$ is  an increasing function for $x\geq 0$, then it is obvious that $\frac{\partial C_{1}(\rho)}{\partial z}\geq 0$. Next,
let us prove this fact.

\[
\begin{aligned}
 f'(x)=&-\frac{1}{x^2}\ln\frac{1+x}{1-x}+\frac{2}{x(1-x^2)}\\
 &=\frac{1}{x^2}\Big(\frac{2x}{1-x^2}-\ln\frac{1+x}{1-x}\Big).
  \end{aligned}
\]

let  $g(x)=\frac{2x}{1-x^2}-\ln\frac{1+x}{1-x}$.
If $x=0$, then $g(x)=0$, and $\forall x\geq 0$, we have
\[
\begin{aligned}
 g'(x)&=\frac{2+2x^2}{(1-x^2)^2}-\frac{2}{1-x^2}=\frac{4x^2}{(1-x^2)^2}\geq 0.
  \end{aligned}
\]
So, $g(x)\geq 0$ for all $x\geq 0$. Therefore, $f'(x)\geq 0$.
According to the above  fact, we can easily know $\frac{\partial C_{1}(\rho)}{\partial z}\geq 0$.

(3) We consider the derivation of $C_{\frac{1}{2}}(\rho)$ with $z$.
$\frac{\partial C_{\frac{1}{2}}(\rho)}{\partial z}=-r_{\frac{1}{2}}^{-\frac{1}{2}}\frac{\partial r}{\partial z}\geq 0$,  the proof of $\frac{\partial r\frac{1}{2}}{\partial z}\leq 0$ will be provided in appendix. Since $\frac{\partial r\frac{1}{2}}{\partial z}\leq 0$, then we have $\frac{\partial C_{\frac{1}{2}}(\rho)}{\partial z}\geq 0$.

 In accordance with the above discussion,  for a fixed $t$, $C_{\alpha}(\rho(t,z))$ have maximum when $z=\sqrt{1-t^2}$, and have  minimum when $z=0$, where $\alpha=2,1,\frac{1}{2}$. Therefore, we consider two special states.

 \begin{equation}
\rho_{max}(t)=\rho(t,\sqrt{1-t^2})
=\left[ \begin {array}{cc} \frac{1+\sqrt{1-t^2}}{2}&\frac{t}{2}\\ \noalign{\medskip}\frac{t}{2}&\frac{1-\sqrt{1-t^2}}{2}\end {array} \right]
 \end{equation}

  \begin{equation}
  \rho_{min}(t)=\rho(t,0)=\left[ \begin {array}{cc} \frac{1}{2}&\frac{t}{2}\\ \noalign{\medskip}\frac{t}{2}&\frac{1}{2}\end {array} \right].
  \end{equation}

 For any single-qubit mixed state $\rho(t,z)$ with the form Eq. (6), $C_{\alpha}(\rho_{min}(t))$ is the lower bound of $C_{\alpha}(\rho(t,z))$, $C_{\alpha}(\rho_{max}(t))$  is the upper bound of $C_{2}(\rho(t,z))$, where $\alpha=2,1,\frac{1}{2}$. Before analyzing the ordering states with $C_{\alpha}$ and $C_{l_{1}}$, we calculate these three Tsallis relative $\alpha$-entropies of coherence of these two special states, where $\alpha=2,1,\frac{1}{2}$. By substituting Eq. (11), Eq. (12) into Eq. (7), Eq. (9), Eq. (10), We get:

\[
\begin{aligned}
C_{2,max}(t)=&C_{2}(\rho_{max}(t))=t,\\
C_{2,min}(t)=&C_{2}(\rho_{min}(t))=t^2. \\
 C_{1,max}(t)=&C_{r}(\rho_{max})\\
 =&-\frac{1+\sqrt{1-t^2}}{2}\ln\frac{1+\sqrt{1-t^2}}{2}\\
 &-\frac{1-\sqrt{1-t^2}}{2}\ln\frac{1-\sqrt{1-t^2}}{2},\\
 C_{1,min}(t)=& C_{r}(\rho_{min})\\
 =&\frac{1+t}{2}\ln\frac{1+t}{2}+
 \frac{1-t}{2}\ln\frac{1-t}{2}+\ln2.\\
C_{\frac{1}{2},max}(t)=&-2\big[(\frac{2-t^2}{2})^\frac{1}{2}-1\big],\\
C_{\frac{1}{2},min}(t)=&-2\big[(\frac{1+\sqrt{1-t^2}}{2})^\frac{1}{2}-1\big].
 \end{aligned}
\]



For any $t\in[0,1]$, and $\alpha=2,1,\frac{1}{2}$, $C_{\alpha}(\rho_{max})$, $C_{\alpha}(\rho_{min})$ are two functions related to variable $t$, and  $l_{1}$ norm of coherence of state $\rho(t,z)$ is  equal to $t$.
These two functions will form a closed region. For any state $\rho(t,z)$ with the form Eq. (6),
$(t,C_{\alpha}(\rho(t,z)))$ will correspond to  a point in closed region. Our main result will
be obtained as the following.

\textbf{Result 2}: $C_{\alpha}$ and $C_{l_{1}}$ don't generate the
same ordering for some single-qubit states with form Eq. (6), where $\alpha=2,1,\frac{1}{2}$.

We will only analyze the ordering states with $C_{2}$ and $C_{l_{1}}$ as presented in Fig. 2. $C_{1}$,  $C_{\frac{1}{2}}$ and $C_{l_{1}}$ are similar as presented in Fig. 3,  Fig. 4.
 Let $\rho(t,z)$ be a single-qubit state  with the form
Eq. (6), and correspond to a point  $(C_{l_{1}(\rho(t,z))}, C_{2}(\rho(t,z)))$ = $(t,C_{2}(\rho(t,z)))$. We can easily
find all states which violate the condition of Eq. (1). If $\rho(t,z)$ correspond to point $O$,
 we can see that $\rho(t,z)$   and any state corresponding to a point in region $OAB$  will  violate the condition of Eq. (1). However, $C_{2}$ and $C_{l_{1}}$  will give the same ordering for $\rho(t,z)$  and any state corresponding to a point outside  region $OAB$.
If a point $Z$ replaces point $O$, then region $OAB$ will be replaced by regions $ZXY$ and $ZMN$. An explicit example will be given as follow. We give two states:

\begin{equation}
\rho_{1}=\left[ \begin {array}{cc} \frac{1}{2}&\frac{1}{4}\\ \noalign{\medskip}\frac{1}{4}&\frac{1}{2}\end {array} \right],
  \rho_{2}=\left[ \begin {array}{cc} \frac{5+\sqrt{21}}{10}&\frac{1}{5}\\ \noalign{\medskip}\frac{1}{5}&\frac{5-\sqrt{21}}{10}\end {array} \right].  \nonumber
  \end{equation}

Substituting $\rho_{1}, \rho_{2}$ into Eq.(7), Eq.(9) and Eq.(10), with an easy calculation, we have
$C_{l_{1}}(\rho_{1})=\frac{1}{2}$,$C_{l_{1}}(\rho_{2})=\frac{2}{5}$,
$C_{2}(\rho_{1})=\frac{1}{4}$, $C_{2}(\rho_{2})=\frac{2}{5}$,
$C_{1}(\rho_{1})\approx 0.13081$, $C_{1}(\rho_{2})\approx 0.17344$,
$C_{\frac{1}{2}}(\rho_{1})\approx 0.0681$, $C_{\frac{1}{2}}(\rho_{2})\approx 0.0817$.
It is clear that $C_{l_{1}}(\rho_{1})\geq C_{l_{1}}(\rho_{2})$ but $C_{\alpha }(\rho_{1})\leq C_{\alpha}(\rho_{2})$,  $\alpha=2,1,\frac{1}{2}$.
It means that $C_{l_{1}}$ and $C_{\alpha}$ generate the different ordering for $\rho_{1}$ and $\rho_{2}$, for any $\alpha=2,1,\frac{1}{2}$.

\begin{center}
\vspace{-0.1cm}
\includegraphics [scale=0.25]{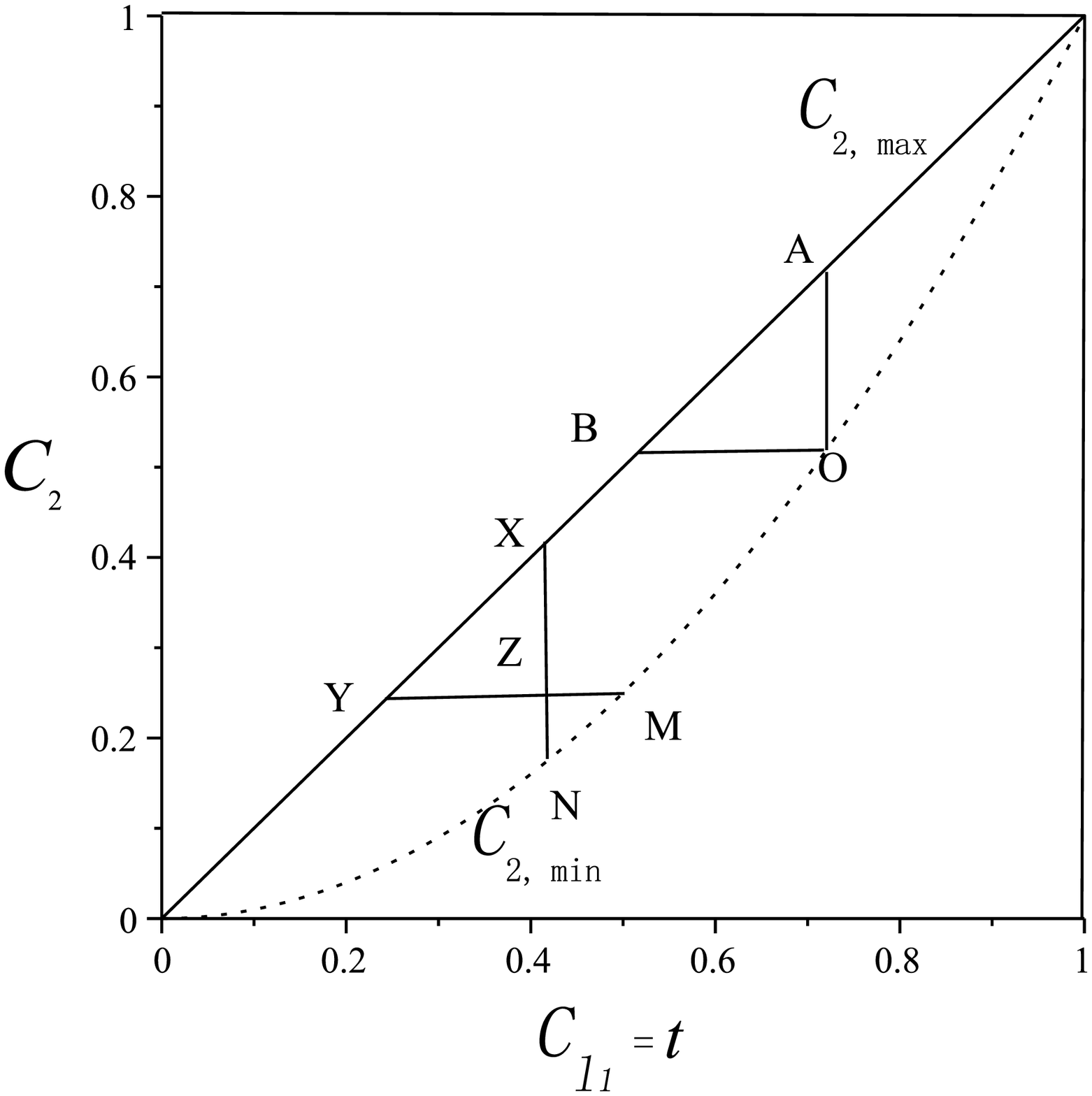}
\vspace{-0.1cm}
\center{Fig. 2. A special Tsallis relative $\alpha$-entropies of coherence $C_{2}$ versus $l_{1}$ norm of coherence $C_{l_{1}}$.}
\end{center}

\begin{center}
\vspace{-0.1cm}
\includegraphics [scale=0.25]{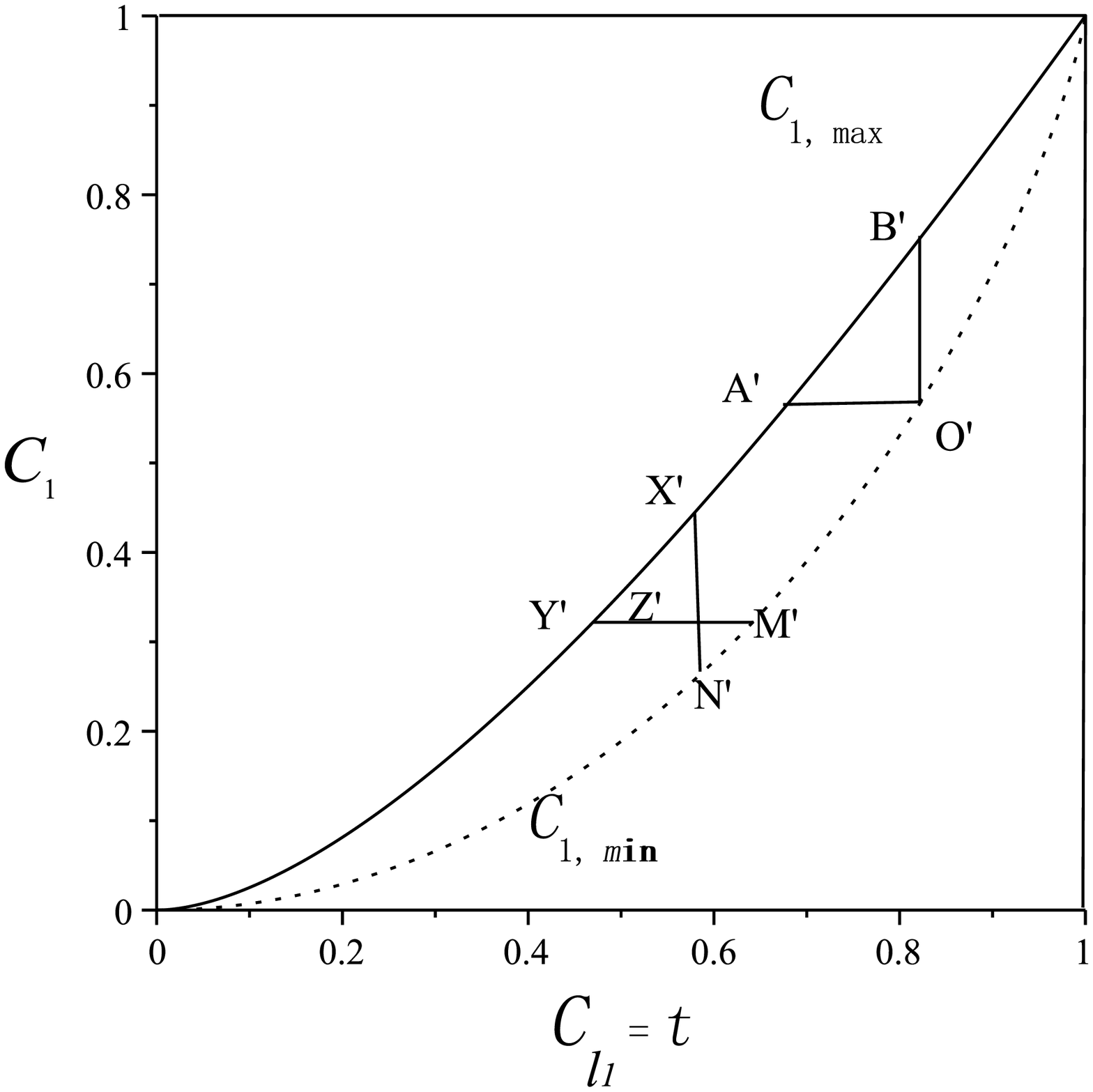}
\vspace{-0.1cm}
\center{Fig. 3. A special Tsallis relative $\alpha$-entropies of coherence $C_{1}$ versus $l_{1}$ norm of coherence  $C_{l_{1}}$.}
\end{center}

\begin{center}
\vspace{-0.1cm}
\includegraphics [scale=0.25]{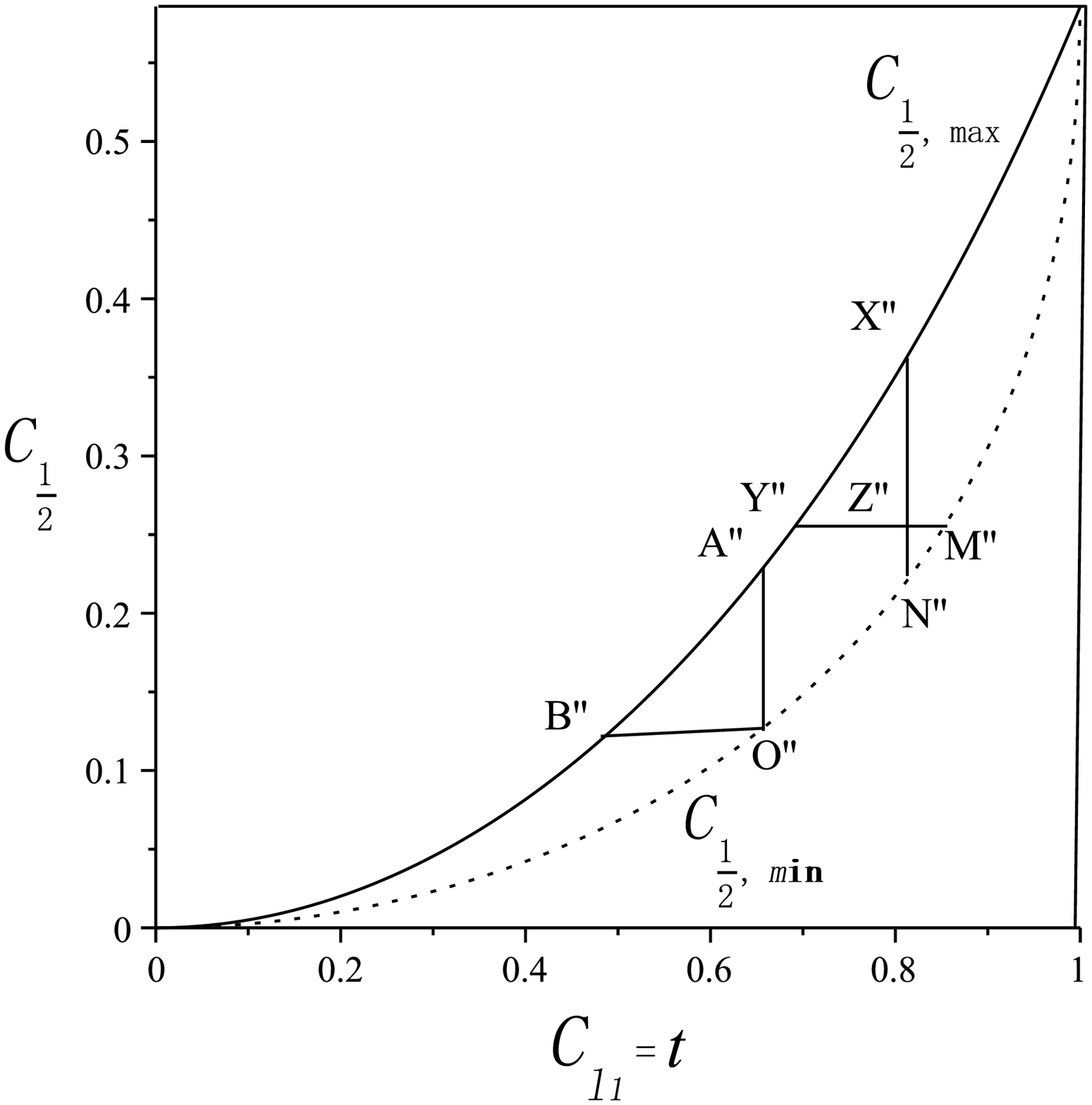}
\vspace{-0.1cm}
\center{Fig. 4. A special Tsallis relative $\alpha$-entropies of coherence $C_{\frac{1}{2}}$ versus $l_{1}$ norm of coherence  $C_{l_{1}}$.}
\end{center}


In the above, we have shown Tsallis relative $\alpha$-entropies of coherence $C_{\alpha}$ and  $l_{1}$ norm of coherence generate different ordering for some single-qubit states when $\alpha$ as some special values, such as 2,1,$\frac{1}{2}$. We conjecture these results remain valid for any $\alpha\in (0,2]$.

\textbf{Conjecture 1}:  For any $\alpha \in (0,2]$, $C_{\alpha}$ and $C_{l_{1}}$ don't generate the same ordering for some single-qubit states with the form Eq. (6).

In Sec.~\ref{sec:pure}, we find  any two Tsallis relative $\alpha$-entropies of coherence $C_{\alpha_{1}}$ and  $C_{\alpha_{2}}$, $\alpha_{1},\alpha_{2}\in (0,2]$, give the same ordering for any single-qubit pure states. Now we consider whether this result is still valid for any single-qubit mixed states. We give  a counterexample to show that it is not true. We give three mixed states with the form Eq. (6).

\begin{equation}
\rho_{1}=\rho(0.5,0.5)=\left[ \begin {array}{cc} 0.75&0.5\\ \noalign{\medskip}0.5&0.25\end {array} \right],\nonumber
\end{equation}

\begin{equation}
\rho_{2}=\rho(0.48,0.58)=\left[ \begin {array}{cc} 0.79&0.24\\ \noalign{\medskip}0.24&0.21\end {array} \right],\nonumber
\end{equation}

\begin{equation}
\rho_{3}=\rho(0.48,0.64)=\left[ \begin {array}{cc} 0.82&0.24\\ \noalign{\medskip}0.24&0.18\end {array} \right]. \nonumber
\end{equation}

By using of Eq. (4), we can have
$C_{1}(\rho_{1})=0.1458$,
$C_{1}(\rho_{2})=0.1400$,
$C_{1}(\rho_{3})=0.1463$,
$C_{2}(\rho_{1})=0.3090$,
$C_{2}(\rho_{2})=0.3100$,
$C_{2}(\rho_{3})=0.3326$,
$C_{\frac{1}{2}}(\rho_{1})=0.0746$,
$C_{\frac{1}{2}}(\rho_{2})=0.0707$,
$C_{\frac{1}{2}}(\rho_{3})=0.0733$.
It is clear that

(1) $C_{1}(\rho_{1})>C_{1}(\rho_{2})$ but $C_{2}(\rho_{1})< C_{2}(\rho_{2})$,

(2) $C_{\frac{1}{2}}(\rho_{1})> C_{\frac{1}{2}}(\rho_{2})$ but $C_{2}(\rho_{1})< C_{2}(\rho_{2})$,

(3) $C_{\frac{1}{2}}(\rho_{1})> C_{\frac{1}{2}}(\rho_{3})$ but $C_{1}(\rho_{1})< C_{1}(\rho_{3})$.

So, we can know that any two of these three Tsallis relative $\alpha$-entropies of coherence don't give the same ordering for some single-qubit mixed states. We conjecture this result is also effective for all Tsallis relative $\alpha$-entropies of coherence.

\textbf{Conjecture 2}:  For any $\alpha_{1},\alpha_{2}  \in (0,2]$, $C_{\alpha_{1}}$ and $C_{\alpha_{2}}$ don't give the same ordering for some single-qubit mixed states with form Eq. (6).

\section{conclusion}\label{sec:conclusion}

In this paper, we studied the ordering with $l_{1}$ norm of coherence measures and Tsallis relative $\alpha$-entropies of coherence for single-qubit states. First, we  showed that any  Tsallis $\alpha$-entropies of coherence  and $l_{1}$ norm of coherence give the same ordering for single-qubit pure states, but this result is not true for  high dimensional pure states, even though these states are pure.
Second, we investigated  some special Tsallis, $\alpha$-entropies of coherence, such as $C_{1}$, $C_{2}$ and $C_{\frac{1}{2}}$. We found any one of these three measures and $C_{l_{1}}$ don't  generate the same ordering for single-qubit mixed states. For any single-qubit state, as presented in Fig. 2, Fig. 3, Fig. 4, we could find all  states which will violate the condition (1). We conjectured  that above results remain valid for any Tsallis relative $\alpha$-entropies of coherence.
Finally, we considered that any two of these three special measures don't generate the same ordering for single-qubit mixed states by a counter-example. Furthermore, we conjectured that it is also true for any two Tsallis relative $\alpha$-entropies of coherence.

\section{acknowledgments}

This paper is supported by NSFC(Grants No. 11271237, No.61228305), 
the Higher School Doctoral Subject Foundation of Ministry of Education of China(Grant No.20130202110001).
\vspace{-0.1cm}
\section{appendix}
\vspace{-0.1cm}
We provide a proof of $\frac{\partial r}{\partial z}\geq 0$. The first equation comes from the derivation of $r_{\frac{1}{2}}$ with respect $z$. In the second equation, we use distributive law and then unite like terms. The last  inequality comes from the fact  $2-z^2-t^2\geq 2\sqrt{1-\sqrt{z^2+t^2}}$.

 \begin{table*}
\[
\begin{aligned}
 \frac{\partial r_{\frac{1}{2}}}{\partial z}=&\frac{1}{\sqrt{2}}[\frac{\sqrt{1+\sqrt{z^2+t^2}}(\sqrt{z^2+t^2}+z)}
 {\sqrt{z^2+t^2}}+\frac{\sqrt{1-\sqrt{z^2+t^2}}(\sqrt{z^2+t^2}-z)}{\sqrt{z^2+t^2}}]\\
 &[\frac{\sqrt{2}}{8}\frac{z(\sqrt{z^2+t^2}+z)}{\sqrt{1+\sqrt{z^2+t^2}(z^2+t^2)}}-
 \frac{\sqrt{2}}{8}\frac{z(\sqrt{z^2+t^2}-z)}{\sqrt{1-\sqrt{z^2+t^2}(z^2+t^2)}}
 -\frac{1}{2\sqrt{2}}\frac{\sqrt{1+\sqrt{z^2+t^2}}z(\sqrt{z^2+t^2}+z)}{(z^2+t^2)^\frac{3}{2}}\\
 +&\frac{1}{2\sqrt{2}} \frac{\sqrt{1+\sqrt{z^2+t^2}}(\sqrt{z^2+t^2}+z)}{(z^2+t^2)}
 -\frac{1}{2\sqrt{2}} \frac{\sqrt{1-\sqrt{z^2+t^2}}z(\sqrt{z^2+t^2}-z)}{(z^2+t^2)^\frac{3}{2}}
 -\frac{1}{2\sqrt{2}} \frac{\sqrt{1-\sqrt{z^2+t^2}}(\sqrt{z^2+t^2}-z)}{(z^2+t^2)}]\\
 +&\frac{1}{\sqrt{2}}[\frac{\sqrt{1+\sqrt{z^2+t^2}}(\sqrt{z^2+t^2}-z)}{\sqrt{z^2+t^2}}+\frac{\sqrt{1-\sqrt{z^2+t^2}}
 (\sqrt{z^2+t^2}+z)}{\sqrt{z^2+t^2}}]\\
 &[\frac{\sqrt{2}}{8}\frac{z(\sqrt{z^2+t^2}-z)}{\sqrt{1+\sqrt{z^2+t^2}(z^2+t^2)}}-
 \frac{\sqrt{2}}{8}\frac{z(\sqrt{z^2+t^2}+z)}{\sqrt{1-\sqrt{z^2+t^2}(z^2+t^2)}}
 -\frac{1}{2\sqrt{2}} \frac{\sqrt{1+\sqrt{z^2+t^2}}(\sqrt{z^2+t^2}-z)}{(z^2+t^2)}\\
-&\frac{1}{2\sqrt{2}}\frac{\sqrt{1+\sqrt{z^2+t^2}}z(\sqrt{z^2+t^2}-z)}{(z^2+t^2)^\frac{3}{2}}
 +\frac{1}{2\sqrt{2}} \frac{\sqrt{1-\sqrt{z^2+t^2}}(\sqrt{z^2+t^2}+z)}{(z^2+t^2)}
 -\frac{1}{2\sqrt{2}} \frac{\sqrt{1-\sqrt{z^2+t^2}}z(\sqrt{z^2+t^2}+z)}{(z^2+t^2)^\frac{3}{2}}].\\
 =&\frac{1}{2}\frac{(\sqrt{z^2+t^2}+z)^2}{(z^2+t^2)^\frac{3}{2}}
 -\frac{1}{2}\frac{(\sqrt{z^2+t^2}-z)^2}{(z^2+t^2)^\frac{3}{2}}
 -\frac{1}{2}\frac{z(\sqrt{z^2+t^2}-z)^2}{(z^2+t^2)^2}
 -\frac{1}{2}\frac{z(\sqrt{z^2+t^2}+z)^2}{(z^2+t^2)^2}\\
 -&\frac{zt^2\sqrt{1-(z^2+t^2)}}{(z^2+t^2)^2}
 +\frac{1}{4}\frac{zt^2\sqrt{1-\sqrt{z^2+t^2}}}{\sqrt{(1+\sqrt{z^2+t^2})(z^2+t^2)^\frac{3}{2}}}
  -\frac{1}{4}\frac{zt^2\sqrt{1+\sqrt{z^2+t^2}}}{\sqrt{1-\sqrt{z^2+t^2}}(z^2+t^2)^\frac{3}{2}}\\
 =&\frac{2z\sqrt{z^2+t^2}}{(z^2+t^2)^\frac{3}{2}}-\frac{z(2z^2+t^2)}{(z^2+t^2)^2}
 -\frac{zt^2\sqrt{1-(z^2+t^2)}}{(z^2+t^2)^2}
 -\frac{1}{2}\frac{zt^2\sqrt{z^2+t^2}}{\sqrt{1-(z^2+t^2)}(z^2+t^2)^\frac{3}{2}}\\
=&\frac{z^2}{2\sqrt{1-(z^2+t^2)}(z^2+t^2)}[-2+z^2+t^2+2\sqrt{1-\sqrt{z^2+t^2}}]\leq 0.
 \end{aligned}
\]

\end{table*}

\end{document}